%% file: eckhardt.tex
\documentstyle[epsf,epsfig,multicol,amsmath,amssymb]{europhys}
\begin{document}
\input euromacr

\newcommand{\beq}{\begin{equation}}  
\newcommand{\eeqn}{\end{equation}}  
\newcommand{\bea}{\begin{eqnarray}}  
\newcommand{\eea}{\end{eqnarray}}  
\title{Wannier threshold law for two electron escape in the presence of an 
external electric field}
\shorttitle{Wannier threshold law with electric field}
\author{Bruno Eckhardt$^1$ and Krzysztof Sacha$^{1,2}$}
\institute{$^1$ Fachbereich Physik, Philipps-Universit\"at Marburg,
	     D-35032 Marburg, Germany \\
	     $^2$ Instytut Fizyki im. Mariana Smoluchowskiego,
	     Uniwersytet Jagiello\'nski, \\
	     ul. Reymonta 4,
	     PL-30-059 Krak\'ow, Poland}
\pacs{
\Pacs{34}{80.Dp}{Atomic excitation and ionization by electron impact}
\Pacs{32}{60.+i}{Zeeman and Stark effects }
\Pacs{31}{25.Jf}{Electron correlation calculations for atoms and ions: excited states}
}
\euro{}{}{}{}
\Date{}
\maketitle
\date{\today}
\maketitle

\begin{abstract}
We consider double ionization of atoms or ions by electron impact 
in the presence of a static electric field. As in Wanniers analysis
of the analogous situation without external field the dynamics
near threshold is dominated by a saddle. With a field the 
saddle lies in a subspace of symmetrically escaping electrons.
Near threshold the classical cross section scales with excess energy $E$ like
$\sigma\sim E^\alpha$, where the exponent $\alpha$ can be determined
from the stability of the saddle and does not depend on the field 
strength. For example, if the remaining ion has charge $Z=2$, the 
exponent is 1.292, significantly different from the 1.056 without the field.
\end{abstract}

The threshold behavior of double electron escape from an atom
or ion was first tackled by Wannier in a famous paper in 1953 
\cite{Wannier,Peterkop,Rau71}. On the basis 
of the classical dynamics of two electrons in an atom he concluded
that on account of the electron repulsion the two escaping electrons
are correlated and that the cross section increases 
with excess energy $E$ like $\sigma(E)\propto E^\alpha$,
 with a non-integer
exponent $\alpha$ that depends on the charge
of the remaining ion. Since the original paper many fine details
of the process have been elucidated in both theory and experiment 
(for recent reviews see \cite{briggs,king,rost}). 

Electron--electron correlations are also important in the time-dependent
process of double ionization in strong laser pulses. Measurements of 
the ion and electron momenta
show that in non-sequential double ionization the escaping electrons
prefer symmetry related motions \cite{weber1,rottke,weber3}. The processes
that are important for this double ionization are a matter of debate,
but when discussed within the rescattering model 
\cite{Corkum,Kulander3} similarities to the Wannier
problem show up \cite{Eckhardt,sacha,sacha3}. 
In the rescattering model one electron is temporarily
ionized by tunneling, but driven back to the atom when the field
changes the phase.
During this rescattering event a highly excited two electron complex
close to the nucleus is formed which then decays towards double
ionization. For the field strengths 
where the characteristics of correlated
electron escape are observed it seems crucial that the external
field does not vanish when the decay takes place; otherwise the
electrons do not have enough energy for double ionization
\cite{Eckhardt,sacha}. In our previous presentations of the 
process in a time-dependent field we argued that the 
electron motion is fast compared to 
the field oscillations so that an adiabatic analysis can be
applied. But it is possible to discuss the process also in the 
presence of a static field, where no adiabatic assumption is
needed, and to push the analogy to the Wannier problem further
by deriving the threshold laws for non-sequential double ionization in the
presence of a static field. This is our aim here.

Wanniers analysis divides into two parts: the identification of the
configuration that leads to double ionization at the threshold
and the determination of the threshold law from the stability
exponents of the fixed point. In Wanniers case the threshold
configuration had two electrons escaping symmetrically on opposite
sides of the nucleus, thus minimizing electron repulsion. The
presence of the field introduces a preference for motion in the 
direction of the field gradient. At the threshold for the process
the energy is equally distributed between the two electrons. 
Furthermore, the distance of the electrons to the nucleus should
be the same for otherwise any difference would be amplified by
electron repulsion and the electrons would not escape simultaneously.
Therefore, in the presence of an electric field the configuration
that corresponds to Wanniers is one where the 
electrons escape along trajectories which are reflection symmetric
with respect to the field axis \cite{Eckhardt,sacha}.

\begin{figure}[htb]
\centering{\epsfig{file=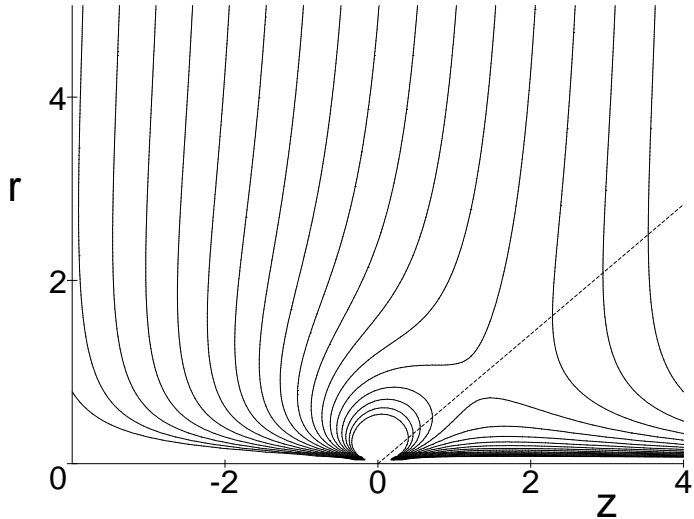,scale=0.4,angle=-90}}
\caption[]{
Potential energy in the symmetric subspace, see Eq.~({\protect \ref{hc2v}}),
for $Z=2$ and $F=1$. When one decreases the amplitude of the field the  
saddle moves along the 
dashed line. 
}\label{equip}
\end{figure}

Due to the rotational symmetry around the field axis
the component of angular momentum along the field axis is conserved 
and for the configuration with lowest threshold this component vanishes.
We can therefore analyze the process in a subspace of zero
angular momentum with symmetrically placed electrons.
In this subspace we introduce cylindrical coordinates 
with $z_1=z_2=z$ the distance along the field direction and 
$\rho_1=\rho_2=r$ the transverse distance. The electrons are on 
opposite sides of the core with respect to the field axis, so that
$\varphi_1=\phi$ and $\varphi_2=\phi+\pi$. The momenta are
$p_{z_1}=p_{z_2}=p_z/2$, $p_{\rho_1}=p_{\rho_2}=p_r/2$ (the factor 2 in the 
momenta is related to the proper canonical transformation and restriction of 
the dynamics to the symmetry subspace \cite{sacha}) 
and $p_{\varphi_1}=p_{\varphi_2}=0$.
With these coordinates for the symmetric subspace the
Hamiltonian becomes, in atomic units,
\begin{equation}
H=\frac{p_r^2+p_z^2}{4}-
\frac{2Z}{\sqrt{r^2+z^2}}
+\frac{1}{2r}
-2Fz \,,
\label{hc2v}
\end{equation}
where $F$ is electric field strength and $Z$ the charge of the
ion after removal
of the two electrons \cite{Eckhardt,sacha}. 
Equipotential curves for $Z=2$ are
shown in Fig.~\ref{equip}. Clearly noticeable is the saddle at
\begin{eqnarray}
r_s=\frac{(2a-1)^{1/4}}{2\sqrt{|F|}}, && 
|z_s|=\frac{(2a-1)^{3/4}}{2\sqrt{|F|}}\,,
\end{eqnarray}
of energy $V_s=-2(2a-1)^{3/4}\sqrt{|F|}$, 
where 
\begin{equation}
a=(2Z^2)^{1/3}.
\label{abr}
\end{equation} 
The ratio $r_s/|z_s|$ depends on the
charge of the nucleus but not on the field strength.
Thus, when the field strength varies the saddle moves along
a line that forms a fixed angle with the field axis,
as indicated in Fig.~\ref{equip}.

The potential shows that in the symmetric subspace 
double ionization requires crossing of the saddle.
In the full phase space this saddle acquires an additional
unstable eigenvalue, so that the symmetric subspace is
an unstable subset of the full phase space. Nevertheless,
it controls double ionization events since trajectories
leading to non-sequential double ionization must
pass close to the saddle in the symmetric subspace: the instability of 
the subspace is connected with an increasing asymmetry in
position, momentum and energy between the two electrons.
Near the threshold for double ionization this 
results in trajectories being pushed towards single
rather than double ionization. 

The stability analysis of the saddle in full configuration space
gives one neutral direction, connected with an overall separable rotation 
around the field axis, three stable directions
and two unstable ones: one unstable direction is the 
`reaction coordinate' \cite{Wigner,Pollak} across the saddle, 
clearly visible in the potential in Fig.~\ref{equip}, 
and the other corresponds to motion away from this subspace.
In the vicinity of the saddle the potential can be expanded
to second order in the deviations from the saddle and the 
Hamiltonian can thus be approximated as
\beq
H\approx V_s+\frac{p_x^2}{2}-\frac{\mu^2x^2}{2}+\frac{p_y^2}{2}-\frac{\nu^2y^2}{2}
+\sum^3_{i=1}\left(\frac{p_{u_i}^2}{2}+\frac{\omega_i^2u_i^2}{2}\right),
\label{harm}
\end{equation}
where $x$ denotes the reaction coordinate in the symmetric subspace,
$y$ the unstable mode away from the
subspace and $u_i$ are the stable 
modes with frequencies $\omega_i$. For the case of zero total angular momentum 
projection on
the field axis we consider here there is no contribution to (\ref{harm}) from
the neutral mode. 
For later reference we note the eigenvalues of the two
unstable directions: for the reaction coordinate it is
\beq
\mu^2=\frac{\sqrt{50 a - 49 + 12/a} - \sqrt{2a-1}}{(2a-1)^{5/4}}F^{3/2}\,,
\label{eigv_reaction}
\end{equation}
and for the motion away from the symmetric subspace it is
\beq
\nu^2=\frac{\sqrt{32 a - 28 + 6/a} + 2\sqrt{2a-1}}{(2a-1)^{5/4}}F^{3/2}\,.
\end{equation}  
The escape from the subspace $y$ is such that $\rho_1=r_s+c_ry$ and 
$z_1=z_s+c_zy$ increase but $\rho_2=r_s-c_ry$ and $z_2=z_s-c_zy$
decrease. The constants $c_r$ and $c_z$ have the same sign and
determine the direction of the unstable mode in the configuration 
space (see \cite{sacha} for further details). 

For the energy equal to the saddle energy only 
a trajectory living in the symmetric
subspace can lead to simultaneous double escape. 
This reduces the dimensionality of the process below that
of the energy shell and consequently 
the cross section vanishes. 
For higher energy some deviations from the
symmetric configuration are allowed, bringing the subset to 
non-zero measure on the energy shell. The
energy dependence of the cross section near threshold is determined by
a competition between the two unstable modes \cite{Rost}.
This is in close correspondence to
the motion on the Wannier ridge where only trajectories 
sufficiently close to 
the symmetric configuration can leave the Coulomb zone and ionize 
\cite{Wannier}.

All trajectories leading to the simultaneous
escape of the electrons have to pass near the saddle. Thus,
to estimate the cross section close to the threshold we may employ the
Hamiltonian of the system in the harmonic approximation, Eq.~(\ref{harm}).
The cross section can be calculated from the phase space flux 
\cite{Wannier,Pollak} associated with the trajectories that lead to 
double ionization. With ${\cal P}$ the projector onto these trajectories,
the microcanonical phase space flux $j_\sigma$ at energy $E$ that crosses 
the saddle can be calculated at the surface $x_0=0$ with the phase 
space velocity projected onto the normal of that surface,
\bea
j_\sigma&=&\int p_{x_0}\,\rho\,\delta(x_0)\,\delta(E-H)\,{\cal P}\,
dp_{x_0}dx_0dp_{y_0}dy_0
\prod_{i=1}^3dp_{u_{i0}}du_{i0} \nonumber\\
&=&\int\, \rho\, {\cal P}\,dp_{y_0}dy_0\,\prod_{i=1}^3 dp_{u_{i0}} du_{i0}\,,
\label{cross}
\eea
where $\rho=\rho(p_{x_0},x_0,p_{y_0},y_0,p_{w_{i0}},w_{i0})$ 
is the initial phase space density.

We are interested in the dependence of 
the cross section on the energy $\varepsilon=E-V_s$ above the saddle. 
The integration limits of the stable degrees of freedom are arbitrary but
finite, so that they cannot go to infinity, as required for the escape
from the reaction zone towards ionization \cite{Wannier}.
The stable directions thus do not contribute to the energy dependence,
just as in Wanniers example \cite{Wannier,Greene,Rost}. Thus the only 
degree of freedom with a critical energy dependence in (\ref{cross})
is the unstable mode associated with $y_0$ and $p_{y_0}$.
Assuming the initial phase
space distribution to be approximately uniform and energy independent
\cite{Wannier} we 
get $j_\sigma(E)\propto\int {\cal P} dp_{y_0}dy_0$.

Taking the initial reaction coordinate on one side of the saddle, 
e.g. $x_0<0$ with $p_{x_0}>0$, the initial conditions of the 
unstable mode must be chosen so that after crossing the saddle 
both electrons escape. Since $x$ and $y$ are center of mass and relative 
coordinates, respectively, double escape requires that both
$x+y$ and $x-y$ simultaneously go to infinity. Thus, the projection
operator ${\cal P}$ selects those orbits that 
at some distance $x$ after crossing the saddle satisfy 
\beq
|y|<|x| \,.
\label{ineq}
\end{equation}
From this it follows that initially only a small 
amount of energy $\varepsilon=E-V_s$ can be put in the $y$ mode
because the Lyapunov exponent for the desymmetrization 
is larger than that for motion along the reaction coordinate, 
$\nu>\mu$. Hence for small $\varepsilon$, 
the initial momentum $p_{x_0}$ can be approximated as
\beq
p_{x_0}=\sqrt{2\varepsilon+\mu^2x_0^2-p_{y_0}^2+\nu^2y_0^2}
\approx \mu |x_0|+\frac{\varepsilon}{\mu |x_0|}.
\label{approx}
\end{equation}
Expressing $y$ in terms of the initial conditions gives, for large time,
\beq
y\approx \frac{1}{2}\left(y_0+\frac{p_{y_0}}{\nu}\right)
\left(\frac{2x}{x_0+p_{x_0}/\mu}\right)^{\nu/\mu}
\label{yapprox}
\end{equation}
Substituting (\ref{yapprox}) and (\ref{approx}) into {(\ref{ineq}) then
results in 
\beq
\left|y_0+\frac{p_{y_0}}{\nu}\right|< \mbox{const} 
\cdot \varepsilon^{\nu/\mu},
\label{restr}
\end{equation}
which is precisely the restriction we need. Changing to canonical variables
$y'=y+p_y/\nu$ and $p_{y'}=p_y/2-\nu y/2$, one finds that
$p_{y'}(t)=p_{y'0}\exp(-\nu t)$ while $y'(t)=y'_0\exp(\nu t)$. 
The inequality (\ref{restr}) is thus the restriction on the 
unstable direction in the phase
space. The initial condition along the stable direction, $p_{y'0}$, 
can be arbitrary -- it is, of course,  
finite due to the requirement that ionizing trajectories must
emerge from the reaction zone \cite{Wannier}. 
The cross section $\sigma(E)$ is proportional to the flux $j_\sigma$, so
that we find for the threshold behaviour the law
\beq
\sigma(E)\propto j_\sigma \propto 
\int {\cal P} \, dy_0' \propto (E-V_s)^\alpha,
\label{power}
\end{equation}
with the exponent $\alpha$ given by
\beq
\alpha^2 = \frac{\nu^2}{\mu^2}=
\frac{\sqrt{32 a - 28 + 6/a} + 2\sqrt{2a-1}}
{\sqrt{50 a - 49 + 12/a} - \sqrt{2a-1}} \,,
\label{expon}
\end{equation}
where $a$ is defined in Eq.~(\ref{abr}). 
In Fig.~\ref{expplot} we show 
$\alpha$ as a function of the nuclear charge together with 
the Wannier exponent; the values are also listed in Table~I. 
For increasing charge $Z$
the exponent $\alpha$ decreases and approaches the limiting value 
$\sqrt{3/2}\approx1.225$.

The exponent $\alpha$ does not depend on the field amplitude but
only on the charge of the nucleus. This raises the question how
the classical Wannier theory can be recovered for vanishing 
field strength $F$.
The answer can be given easily by appealing to Wanniers analysis: he 
introduced a Coulomb dominated zone inside which potential energy exceeds 
kinetic energy. The outer boundary of this zone is then energy dependent
and increases without bound as the total energy approaches zero.
As long as the saddle induced by the field is inside this radius
we expect the exponent given above (\ref{expon}), but if it is
outside we expect the classical Wannier exponent. In the limit of
vanishing field strength but for fixed excess energy the saddle moves 
outside the Coulomb zone and the classical exponents are restored.
This transition might be accessible experimentally.

\begin{figure}[htb]
\centering{\epsfig{file=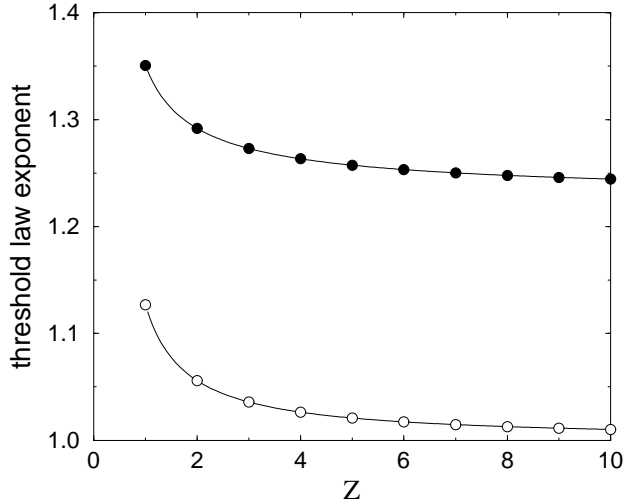,scale=0.4,angle=-90}}
\caption[]{Exponents in the threshold law for different charges 
of the nucleus.
Full circles correspond to Eq.~({\protect \ref{expon}}) in the presence
of an external field while open symbols give the  
Wannier prediction without field {\protect \cite{Wannier}}. 
}\label{expplot}
\end{figure}

The classical Wannier theory for double ionization without external field
gives the same threshold law as do 
semiclassical and quantum calculations \cite{Peterkop,Rau71,rost}. 
This may be traced to the fact that a remarkable scaling law relates
the limits of energy approaching threshold and of Planck's
constant becoming small \cite{Wannier,rost}. In the present case
the saddle is at a finite distance and can be overcome by tunneling.
This will modify the threshold law very close to the classical saddle, 
in an energy interval of about $\hbar\mu$, with $\mu$ the frequency
for motion along the reaction coordinate (eq. (5)). In the semiclassical
limit and for weak fields this can be made arbitrarily small, so
that the algebraic behaviour should become accessible. 

The most direct experimental study of the proposed threshold behavior
would be double-photoionization \cite{briggs,king}
in the presence of an external field. The photon resolution
of 0.1~eV achieved in recent experiments \cite{Huetz} should enable
detection of deviations from Wannier theory e.g. in double-photoionization
of He atoms in a static field of 
30~kV/cm, where the saddle energy is $-0.3$~eV.

Pulsed lasers provide stronger fields but add a time-dependence. 
However, the laser field can be considered as stationary if the time
needed to cross the barrier is a small fraction of the field cycle, only. 
Double ionization could be triggered by a crossed electron beam.
The energy of the saddle and of the electronic beam change with the 
field phase but both are well defined. The field intensity can then
be adjusted so that the excess energy reaches the energy of the saddle
only at specific moments in time. Such an experiment would also provide
a very interesting direct test for the rescattering model for
double ionization in strong fields \cite{Corkum,Kulander3}. 

Finally we should mention that the correlated escape used to derive
the threshold law is not the only pathway to double ionization. 
Trajectory studies \cite{Eckhardt,sacha} show that initial conditions
started near the saddle that do not lead to immediate double
ionization can lead to sequential ionization in that one electron 
escapes immediately but the other escapes after a return to the nucleus.
However, in such a sequential escape the momenta of the second electron
are not correlated to the ones of the first electron and this
can perhaps be used to
distinguish correlated from sequential escape \cite{sacha}.

To conclude we have presented the analysis of the threshold law for the double
ionization in electron impact in the presence of a static electric field.
Simultaneous ionization of the electrons then proceeds via a correlated 
crossing of the saddle induced by the external field. 
The cross section behavior close to the
threshold is algebraic in excess energy with an exponent determined from
the ratio of the positive Lyapunov exponents of 
the unstable modes of the saddle. We hope that the generalization of
the Wannier theory to the case with an additional external field presented
here will stimulate 
experimental and further theoretical investigations. 
\\

Financial support by the Alexander von Humboldt
Foundation is gratefully acknowledged. 
The work is also a part of KBN project No. 5~P03B~088~21 (K.S.).

\begin{table}
\renewcommand{\arraystretch}{1.1}
\begin{tabular}{lcc}
\hline
$Z$ & $\alpha$ & Wannier exponent \\
\hline
1 & 1.351 & 1.127 \\
2 & 1.292 & 1.056 \\
3 & 1.273 & 1.036 \\
4 & 1.263 & 1.026 \\
5 & 1.257 & 1.021 \\
\hline
\end{tabular}
\caption[]{Threshold exponent $\alpha$, Eq.~(\ref{expon}), 
and  Wannier exponent for different charge of the remaining ion.}
\end{table}

\end{document}

%% file: euromacr.tex
\def\macro#1{$\backslash${\tt #1}}

\def\etal{{\hbox{{\tenit\ et al.\/}\tenrm :\ }}}

\def\And{{\rm and\ }}

\def\dfrac#1#2{{\displaystyle\frac{#1}{#2}}} 

\def\drm{{\rm d}}

\def\dif#1#2{\frac{\drm#1}{\drm#2}}

\def\tg{\mathop{\rm tg}\nolimits}
\def\ctg{\mathop{\rm ctg}\nolimits}
\def\arctg{\mathop{\rm arctg}\nolimits}
\def\sec{\mathop{\rm sec}\nolimits}
\def\cosec{\mathop{\rm cosec}\nolimits}
\def\tgh{\mathop{\rm tgh}\nolimits}
\def\ctgh{\mathop{\rm ctgh}\nolimits}
\def\order{\mathop{\rm O}\nolimits}
\def\tr{\mathop{\rm tr}\nolimits}
\def\Div{\mathop{\rm div}\nolimits}
\def\Rot{\mathop{\rm rot}\nolimits}
\def\Grad{\mathop{\rm grad}\nolimits}

\def\centerinsert#1#2#3#4{%
\centerline{\hbox{\vbox to #3cm{\parindent=0pt
\hrule width #4cm height 0pt depth 0pt\vfill%
\special{illustration #1 scaled #2}\medskip}}}}

\def\stars{\bigskip\centerline{***}\medskip}

\newif\ifboo \boofalse

\def\Review#1{\boofalse{\it #1},}
\def\Name#1{{\sc #1},}
\def\Vol#1{\ifboo Vol. {\bf #1}\else{\bf #1}\fi}
\def\Year#1{\ifboo #1\else(#1)\fi}
\def\Book#1{\bootrue{\it #1},}
\def\Page#1{\ifboo {\rm p. #1}\else{\rm #1}\fi}

%% file: eckhardt.bbl
\begin{thebibliography}{99}

\bibitem{Wannier}  
Wannier, G.H., {\it Phys. Rev.}, {\bf 90} (1953) 817.

\bibitem{Peterkop} Peterkop, R., {\it J. Phys. B: At. Mol. Phys.}, {\bf 4}
(1971) 513.

\bibitem{Rau71} Rau, A. R. P., {\it Phys. Rev. A}, {\bf 4} (1971) 207.

\bibitem{briggs} Briggs, J. S. and Schmidt, V.,  
{\it J. Phys. B: At. Mol. Phys.}, {\bf 33} (2000) R1.

\bibitem{king} King, G. C. and Avaldi, L., 
{\it J. Phys. B: At. Mol. Phys.}, {\bf 33} (2000) R215.

\bibitem{rost} Rost, J.M., Phys. Rep., {\bf 297} (1998) 271 

\bibitem{weber1}  Weber, Th.,
Weckenbrock, M., Staudte, A., Spielberger, L., 
Jagutzki, O., Mergel, V., Afaneh, F., Urbasch, G., Vollmer, M., 
Giessen, H. and D\"orner, R., 
{\it Phys. Rev. Lett.}, {\bf 84} (2000) 443.

\bibitem{rottke}
Moshammer, R., 
Feuerstein, B., Schmitt, W., Dorn, A., Schr\"oter, C.D., 
Ullrich, J., Rottke, H., Trump, C., Wittmann, M., Korn, G., Hoffmann, K. 
and Sandner, W.,
{\it Phys. Rev. Lett.}, {\bf 84} (2000) 447.

\bibitem{weber3} Weber, Th., 
Giessen, H., Weckenbrock, M., Urbasch, G., 
Staudte, A., Spielberger, L., 
Jagutzki, O., Mergel, V., Vollmer, M. and D\"orner, R.,
{\it Nature}, {\bf 405} (2000) 658. 

\bibitem{Corkum} Corkum, P.B., {\it Phys. Rev. Lett.}, {\bf 71} (1993)
1994. 

\bibitem{Kulander3} Kulander, K.C., Schafer  K.J. and Krause J.L. 
in {\it Super-Intense Laser-Atom Physics}, Proceedings of the NATO Advanced 
Research Workshop, in Han-sur-Lesse, Belgium, 1993, edited by 
Piraux, B., L'Huillier, A. and Rz\c{a}\.{z}ewski K. (Plenum, New York,
1993), p.~95.

\bibitem{Eckhardt} Eckhardt, B. and Sacha, K., Physica Scripta, {\bf T90} 
(2001) 185.

\bibitem{sacha} Sacha, K. and Eckhardt, B., Phys. Rev. A {\bf 63} (2001) 
043414. 

\bibitem{sacha3} Sacha, K. and Eckhardt, B., Phys. Rev. A, to appear.

\bibitem{Wigner} Wigner, E. P.,
    {\it Z. Phys. Chemie B}, {\bf 19} (1932) 203.

\bibitem{Pollak} E. Pollak, in {\em Theory of Chemical Reactions}, 
	{vol III, M. Baer, ed., CRC Press, Boca Raton, 1985}, p. 123.

\bibitem{Rost} J.M. Rost, Physica E, {\bf 9} (2001) 467.

\bibitem{Greene} Greene, C.H. and Rau, A. R. P., 
{\it Phys. Rev. Lett.}, {\bf 48} 533 (1982).
 
\bibitem{Huetz} Huetz, A. and Mazeau, J., {\it Phys. Rev. Lett.}, 
{\bf 85} (2000) 530.

\end{thebibliography}
